    \documentclass[lettersize,journal]{IEEEtran}
    \usepackage{booktabs}
    \usepackage{diagbox}
    \usepackage{makecell}
    \usepackage{graphicx,amssymb,amsmath}
    \usepackage{multicol}
    \usepackage[noadjust]{cite} 
    \usepackage{setspace}
    \usepackage{subfigure}
    \usepackage{graphicx}
    \usepackage{float}
    \usepackage {url}
    \usepackage{stfloats}
    \usepackage{amsthm,pifont}
    \usepackage{flushend}
    \usepackage{cases,subeqnarray}
    \usepackage{bm,multirow,bigstrut}
    \usepackage{amsmath, amsthm, amssymb}
    \usepackage{textcomp}
    \usepackage{latexsym,bm}
    \usepackage{booktabs}
    \usepackage{xcolor}
    \usepackage{mathtools}
    \usepackage{dsfont}
    \usepackage{extarrows}
    \usepackage{epsfig}
    \usepackage{epsfig}
    \usepackage{epstopdf}
   \usepackage{xcolor}
  \usepackage{array} 
    \usepackage{hyperref}
    
    \theoremstyle{plain}

    \theoremstyle{plain}

    \usepackage{amsmath}

    \IEEEoverridecommandlockouts

\begin{document}
\title{Toward Scalable Generative AI via Mixture of Experts in Mobile Edge Networks
}
\author{Jiacheng Wang, Hongyang Du, Dusit~Niyato,~\IEEEmembership{Fellow,~IEEE}, Jiawen Kang, Zehui Xiong, \\ Dong In Kim,~\IEEEmembership{Fellow,~IEEE}, and Khaled B. Letaief,~\IEEEmembership{Fellow,~IEEE}

\thanks{J.~Wang, H.~Du and D. Niyato are with the School of Computer Science and Engineering, Nanyang Technological University, Singapore (e-mail: jiacheng.wang@ntu.edu.sg, hongyang001@e.ntu.edu.sg, dniyato@ntu.edu.sg).}
  
\thanks{J. Kang is with the School of Automation, Guangdong University of Technology, China. (e-mail: kavinkang@gdut.edu.cn).}
 
\thanks{Z. Xiong is with the Pillar of Information Systems Technology and Design, Singapore University of Technology and Design, Singapore (e-mail: zehui\_xiong@sutd.edu.sg).}

\thanks{D.~I.~Kim is with the Department of Electrical and Computer Engineering, Sungkyunkwan University, Suwon 16419, South Korea (email:dikim@skku.ac.kr).}

\thanks{Khaled B. Letaief  is with the Department of Electrical and Computer Engineering, Hong Kong University of Science and Technology (HKUST), Hong Kong (e-mail: eekhaled@ust.hk).}

\thanks{Corresponding author: Dong In Kim}


}

\maketitle
\begin{abstract}
The advancement of generative artificial intelligence (GAI) has driven revolutionary applications like ChatGPT. The widespread of these applications relies on the mixture of experts (MoE), which contains multiple experts and selectively engages them for each task to lower operation costs while maintaining performance. Despite MoE, GAI faces challenges in resource consumption when deployed on user devices. This paper proposes mobile edge networks supported MoE-based GAI. We first review the MoE from traditional AI and GAI perspectives, including structure, principles, and applications. We then propose a framework that transfers subtasks to devices in mobile edge networks, aiding GAI model operation on user devices. We discuss challenges in this process and introduce a deep reinforcement learning based algorithm to select edge devices for subtask execution. Experimental results will show that our framework not only facilitates GAI’s deployment on resource-limited devices but also generates higher-quality content compared to methods without edge network support.
\end{abstract}

\IEEEpeerreviewmaketitle
\section{Introduction}

Recently, generative artificial intelligence (GAI) based applications, such as ChatGPT and BERT, have gained significant attention. These GAI based large language models (LLM)~\cite{liu2023optimizing} are renowned for their capacity to understand and generate text akin to human speech, significantly advancing various applications that rely on natural language processing~\cite{shen2024large}. For example, ChatGPT, known for its ability to generate and polish text, quickly became part of numerous applications and interacted with millions of users within months of its launch. This rapid integration and widespread use underscore the increasing fascination and reliance on sophisticated artificial intelligence (AI) for complex language tasks.

The rising demand for AI-generated content (AIGC) prompts a review of GAI architecture, focusing on scalability and efficiency. Traditionally, enhancing GAI efficiency involves enlarging the models with more layers and parameters for deeper analysis~\cite{hu2023survey}, but this leads to increased resource consumption and impacts the model's scalability. For instance, GPT-4 contains approximately 1.76 trillion parameters\footnote{https://openai.com/research/gpt-4}, can only be deployed in cloud-based environments and necessitates a substantial amount of resources for both training and inference. Therefore, blindly expanding model size to support applications is impractical. 

To effectively tackle the aforementioned challenges, integrating the Mixture-of-Experts (MoE) framework into large GAI-based models, such as ChatGPT\footnote{https://openai.com/research/techniques-for-training-large-neural-networks.}, offers a promising approach. Specifically, MoE is an architecture that integrates a set of specialized neural network components, i.e., experts, to handle the given tasks, with each expert fine-tuned to handle specific types of subtask or subdataset~\cite{yuksel2012twenty}. Unlike conventional models that use all parameters for every input, MoE selectively engages only a relevant subset of these experts depending on the input. For example, in Raphael~\cite{xue2023raphael}, the gating network first splits an input sentence (i.e., prompt) into individual words and assigns these words to relevant experts. Then, each expert depicts a particular textual concept onto a specified image region at a diffusion timestep. At last, the gating network integrates the outputs from all experts to generate the final output. This selective activation reduces the overall workload of the model and allows each expert to hone its specialized skills, leading to enhanced performance. 

By integrating the MoE architecture~\cite{masoudnia2014mixture}, GAI-based large models can continue to expand in scale, capabilities, and flexibility. However, further efforts are required to address the deployment of such models in mobile edge networks, which is crucial for providing users with ubiquitous advanced AI capabilities. Deploying MoE-enhanced foundation models in mobile edge networks offers major advantages.

\begin{itemize}
\item \textbf{Reduced Latency}: Distributing subtasks to edge devices for computation enables GAI models for reasoning even with constrained local resources, significantly reducing response latency, which is vital for real-time applications such as virtual assistants or real-time translation.

\item \textbf{Enhanced Privacy and Security}: By distributing non-privacy-threatening subtasks to the selected edge experts in mobile networks and processing sensitive data on user devices, the system circumvents the risk of data leakage, thereby enhancing the overall security of the system.

\item \textbf{Improved Availability}: By distributing computing subtasks among user devices and edge experts, resource usage is spread across multiple devices, lowering the burden on the user device. This enables the deployment of large GAI models on mobile devices, significantly enhancing their availability.
\end{itemize}


However, deploying large models in mobile edge networks presents distinct challenges. The primary one is ensuring that GAI-based large models operate effectively. Therefore, we propose a framework that allows utilizing different devices within the edge networks to assist users in completing AIGC tasks. Unlike existing works assuming sufficient local computing resources, in this paper, the overall resources of the user devices are limited. Therefore, the proposed framework appropriately transfers certain subtasks to other devices in the mobile edge network, considering factors including communication and computing costs. This guarantees smooth operation of GAI models under various resource conditions and workloads. The contributions of this paper are summarized as follows.

\begin{itemize}
\item We comprehensively analyze MoE's structure, principles, and advantages, and then explore its applications. These include not only common areas such as LLM and image generation, but also those in wireless communications, such as signal detection.

\item We propose a novel framework that, when user devices' resources are insufficient, can selectively transfer certain subtasks to edge devices for completion, taking into account factors such as communication and energy consumption costs.

\item Through a case study, we demonstrate that our framework can select appropriate edge devices to handle specific subtasks, ensuring the GAI operates efficiently and achieves service quality close to the upper bound when user device has the limited resources.
\end{itemize}

\section{Generative AI and Mixture of Experts }
\subsection{Generative AI}
Generative AI can emulate human understanding and creativity to generate new digital content. GAI utilizes models such as neural networks to analyze and capture the latent patterns of the training data. Then, the learned patterns are used to generate new outputs~\cite{wang2023guiding}, which is different from the discriminative AI (DAI) that relies on predefined algorithms. This makes GAI excel in data generation, offering several advantages over DAI.

\begin{itemize}
\item \textbf{Enhanced Creativity}: GAI can produce new content, such as text and images, going beyond the DAI, which primarily analyzes and interprets data. 

\item \textbf{Better Personalization}: GAI excels in creating personalized content, achieving a level of personalization beyond DAI's capabilities. 

\item \textbf{Improved Privacy}: GAI can generate anonymized datasets mirroring real data's statistical properties, enabling research while protecting sensitive information.
\end{itemize}

However, enhanced capabilities imply more complex computations and larger model sizes. For instance, the text-to-image models DALL-E and GLIDE possess 12 billion and 3.5 billion parameters, respectively. This triggers the following deployment and application challenges.
\begin{itemize}
\item \textbf{High Operation Costs}: Due to the numerous parameters, the operation of GAI models requires significant energy. Particularly, to enhance their ability to generate better outputs, they need to be trained with large datasets, which significantly raises the operating costs.

\item \textbf{Significant Latency}: During service provision, GAI models need complex computations based on the learned pattern to generate content, consuming significant time and therefore struggling to meet applications with low latency requirements.

\item \textbf{Limited Adaptability}: The size of GAI models makes it challenging to train them to generate new types of content solely through fine-tuning. This indicates that to meet the evolving user demands, they often need to be retrained, making the maintenance and updating difficult.
\end{itemize}

Therefore, it is essential to introduce MoE and task allocation architecture when deploying GAI in mobile edge networks.


\subsection{Mixture of Experts}
In traditional machine learning, models such as supporting vector machines (SVMs) and neural networks, process data uniformly. While effective, this is not optimal, as different models vary in efficiency with different data distributions~\cite{masoudnia2014mixture}. To fully leverage strengths and mitigate weaknesses of each model in applications, the MoE has been proposed. Based on the divide-and-conquer principle, MoE features a tree-like structure with multiple expert units and a gating network. Concretely, in MoE, each expert is trained for high efficiency in its designated domain. Hence, these experts can address specific subtasks or data subsets within the larger problem space. During operation, the gating network acts as the central coordinator, dynamically evaluating the input and deciding which experts are best suited to process it. Moreover, the gating network allocates weights to the experts' outputs, indicating each expert's contribution to the final solution. Given this structure of MoE, the operation of MoE includes the following major steps.


\begin{enumerate}
    \item MoE begins by decomposing a complex task into smaller, more manageable tasks through the gating network. This decomposition is not according to predetermined rules but is adaptive based on the nature of the input data.
    \item The gating network then evaluates and selects the most suitable expert(s) for the each subtask. This selection is not binary but rather involves assigning varying levels of importance to the advice of different experts, allowing for a nuanced combination of their outputs.
    \item The selected experts handle the assigned tasks. Each expert, through its specialized training, develops a deep understanding of its segment of the problem. This specialization enables the expert to efficiently process and respond to the data it is responsible for.
    \item The gating network, after weighing the contributions of each expert, integrates their outputs. This integration is not merely a summation but a complement that considers the relevance and importance of each expert's output.
\end{enumerate}

\begin{figure*}[t]
	\centering
	\includegraphics[width=1\textwidth]{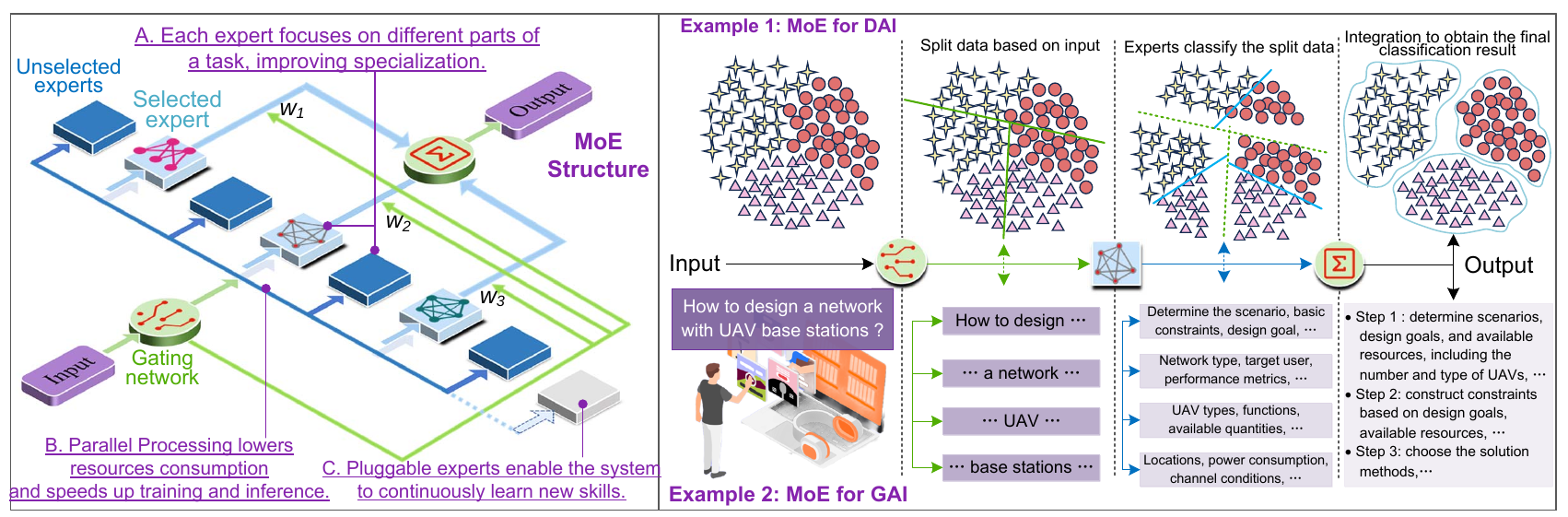}%
	\caption{The structure of MoE, its advantages, and application examples in DAI and GAI. In the application examples, we illustrate the working principles of MoE through data classification (corresponding to DAI) and text generation (corresponding to GAI). }
	\label{MOE}
\end{figure*}
Figure 1 presents the architecture of the MoE and explains its operation through two examples. In the first example, we consider in nonlinear classification where given the input data, the gating function first analyzes the data and divides it into several sub-datasets. Then, each sub-dataset is classified by a selected expert. Finally, the multiple results are aggregated to obtain the final classification result. In this process, the MoE simplifies the nonlinear problem into multiple linear classification ones, which are then handled by models adept in binary classification, thereby enhancing the efficiency. In the second example about LLM, after the user inputs the prompt, the gating network semantically splits it into several parts, e.g., `How to design' and `a network'. After that, different phrases are processed by different experts, generating more detailed information. For instance, experts dealing with `UAV' would consider various aspects of a UAV as base stations and generate comprehensive content, such as UAV types, and available quantities, etc. Subsequently, these are integrated with the results from other experts, such as the type of network and number of users, thereby formulating the final specific steps for network design.

From the above examples, we can see that MoE can not only effectively improve the efficiency of GAI-based large models but also enhance their performance, making them applicable to a wider range of scenarios. Specifically, MoE endows GAI models with several advantages.

\begin{itemize}
\item \textbf{Enhanced Specialization}: MoE uses multiple experts, each specialized in different aspects of a task. In GAI based large models, such specialization allows for a deeper and more focused analysis of different parts of the given prompt than a single algorithm based approach, thereby achieving better performance. 

\item \textbf{Parallel Processing}: MoE allows for parallel training and inference, since each expert operates independently. For large GAI models, this capability not only effectively reduces the consumption of computing resources but also accelerating the training and inference processes.

\item \textbf{Lifelong Learning}: Each expert can continuously learn and enhance its performance over time. Moreover, new experts can be trained and incorporated into the MoE network, along with the expansion of the gating network. This ensures that the GAI models can learn new skills while retaining their original knowledge, thereby meeting the ever-changing demands of the user.
\end{itemize}

This overcomes the challenges aforementioned faced by GAI during deployment and application, establishing a solid foundation for ubiquitous GAI services.

\section{Applications of MoE}
MoE can enhance both DAI and GAI, thereby holding potential for applications across various fields. In this section, we provide examples to illustrate how current systems employ MoE to enhance their performance.
\begin{figure*}[t]
	\centering
	\includegraphics[width=1\textwidth]{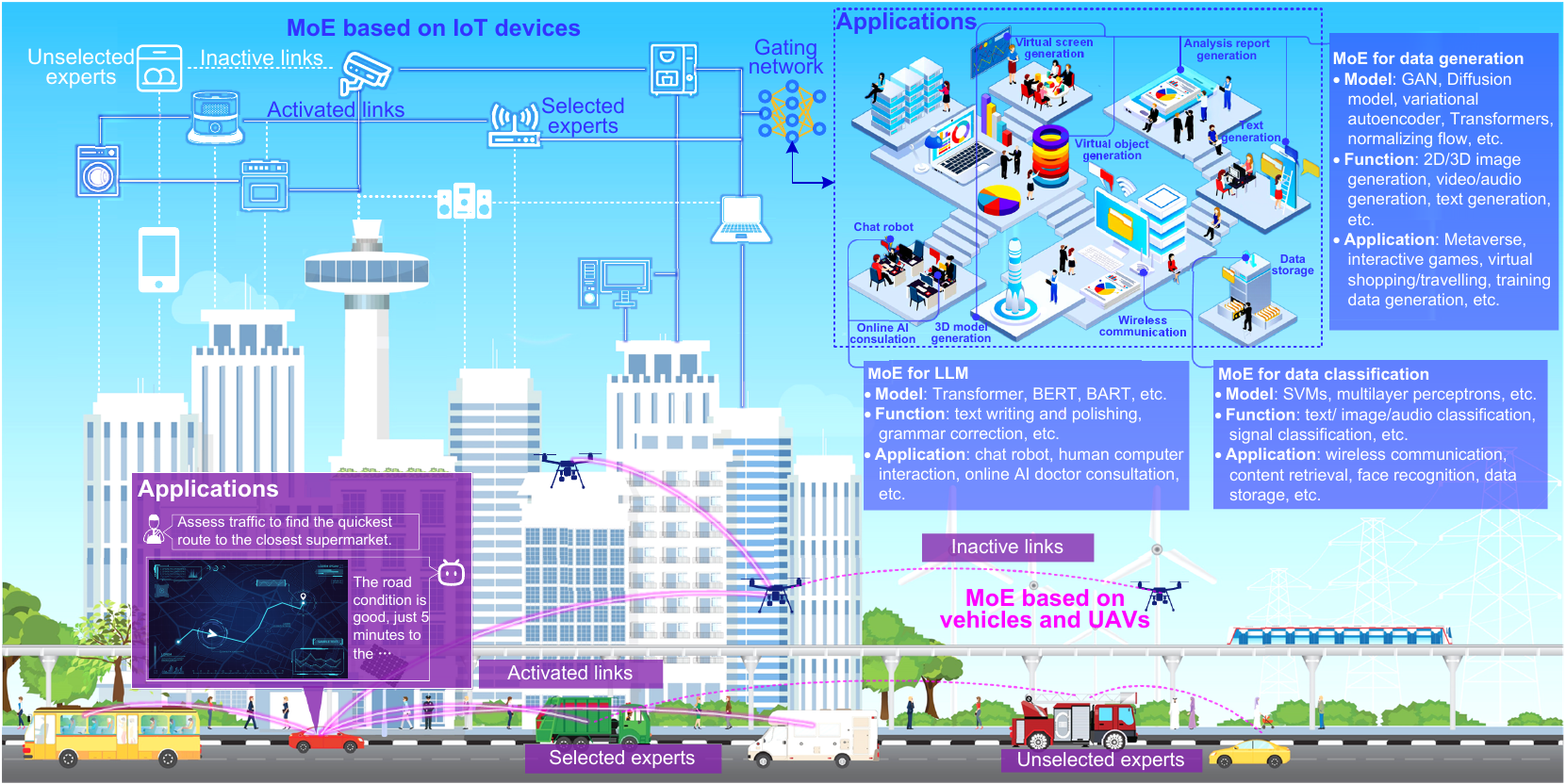}%
	\caption{The applications of MoE. In this figure, activated links correspond to experts selected by the gating function, utilized for the transmission of tasks and generated content. Inactive links correspond to experts that are not selected and hence do not participate in content generation. In DAI, the MoE can be integrated with models such as multilayer perceptrons and SVMs to enhance the accuracy of classifying images, audio, and even wireless signals, thereby supporting applications such as wireless communication, facial recognition, and more. In GAI, MoE can be combined with Diffusion models and Transformers, so as to improve the quality of generated content, including text, 3D images, audio, and video, thereby enabling applications like chatbots, the metaverse, and among others.}
	\label{APLCT}
\end{figure*}
\subsection{Application of MoE-Enhanced DAI Models}

\subsubsection{MoE in Wireless Communications} MoE can be used to enhance the signal classification. For instance, MoE-AMC~\cite{gao2023moe} integrates a multilayer perceptron-based gating network and different experts to realize the automatic modulation classification (AMC). Specifically, one expert is the ResNet based High SNR Recognition Model (HSRM), adept at handling high SNR signals due to its perception and feature abstraction capabilities. The other one is Transformer based Low SNR Recognition Model (LSRM), which excels with low SNR signals by focusing on suppressing noise-induced variations. During operation, the gating network analyzes the input signal, and then directs the high and low SNR signals to the HSRM and LSRM for processing, respectively. The final result combines outputs from HSRM and LSRM, weighted according to the probability of the signal being in the high or low SNR category. Comprehensive evaluation reveals that MoE-AMC achieves a classification accuracy of 95\%, better than 85\% of conventional single-model based systems. In low SNR conditions, it maintains a classification accuracy of around 90\%, exceeding 70-75\% of regular models in similar scenarios. This demonstrates that MoE effectively enhances the performance of AMC, providing robust support for wireless communications.

\subsubsection{MoE in Wireless Sensing} Besides AMC, MoE can also support wireless sensing. In~\cite{zhang2018crosssense}, CrossSense is introduced to boost the Wi-Fi sensing in cross-scenario and large-scale applications through the integration of multiple classifiers and signal features. CrossSense includes a K-Nearest Neighbour (KNN) based gating network and several classifier based experts. During working, the gating network matches the input signal features with the database, identifying the three closest fingerprints to the input. CrossSense then employs the most effective experts for these fingerprints to classify the input and integrates their results to obtain the final classification result. Note that for new environments, CrossSense employs synthetic training samples, generated by its roaming model, to fine-tune the experts, thereby enhancing the experts’ adaptation. Evaluation with 1.2 million samples demonstrate that CrossSense improves the accuracy of WiFi sensing from about 20\% to over 80\% and achieves over 90\% accuracy for CSI-based gesture recognition. Moreover, it maintains consistent performance across problem sizes, with a less than 5\% reduction in accuracy as the problem size increases, better than other schemes that can drop by 50\%.

\subsection{Application of MoE-Enhanced GAI Models}

\subsubsection{MoE in Large Language Models} A representative of LLMs is GLaM~\cite{du2022glam}, which is an integration of the MoE and Transformer. Specifically, GLaM alternates between standard Transformer and MoE layers. The MoE layers consist of multiple feed-forward networks managed by the gating network, which selects the most suitable experts for each token. As an input sequence is fed into GLaM, it traverses both standard Transformer and MoE layers. In the standard layers, the Transformer processes the input, handling the sequential nature of language. In the MoE layers, the gating network selects relevant experts to process each token, leading to a more efficient computation. Subsequently, the resutls from selected experts are integrated through a weighted average to form the output for each token. This process is repeated for every token in the sequence, and hence, the final output can benefit from both Transformer and MoE layers. Despite 1.2 trillion parameters, GLaM consumes only one-third of the energy and requires half the computation flops for inference compared to GPT-3. Meanwhile, with 8\% of its parameters activated per input, GLaM outperforms GPT-3 in 6 out of 7 tasks. Furthermore, GLaM trained with 280 billion tokens notably outperforms GPT-3 trained with 300 billion tokens in multiple learning settings, demonstrating its superior understanding and generation capabilities.

\subsubsection{MoE in Image Generation} In addition to LLM, MoE can support image generation, and RAPHAEL~\cite{xue2023raphael} is one of them. RAPHAEL contains the diffusion model with MoE layers, which are divided into space-MoE and time-MoE layers. Besides, RAPHAEL includes a text gating network that determines the expert responsible for a particular region of the image based on the given prompt. During image generation, RAPHAEL first analyzes the textual and visual content via self-attention and cross-attention layer. Then, the space-MoE layer maps text concepts to the specific image regions using specialized experts. Concurrently, the time-MoE layer controls the image's visual evolution across different timesteps. Finally, the gating network combines the output from these experts to produce the final image of the text prompt. The evaluations based on the MS-COCO 256x256 dataset\footnote{https:cocodataset.org/overview} illustrate that RAPHAEL achieves a ground-breaking zero-shot Frechet Inception Distance (FID-30k) score of 6.61, better than DALL-E 2's 10.39, Stable Diffusion's 8.32, and Imagen's 7.27. This highlights RAPHAEL's superior ability to produce accurate and visually appealing images in response to text prompts compared to its predecessors.

\subsection{Lessons Learned}
From the above applications, we can draw the following key conclusions.
\begin{itemize}
\item MoE can provide support for both GAI and DAI, primarily in three aspects: reducing the system's operation costs, improving the overall performance, and enhancing the scalability.

\item For DAI, the improvements in system performance and scalability are particularly notable. For instance, with the help of MoE, the signal classification accuracy of MoE-AMC~\cite{gao2023moe} can reach up to 90\%, significantly surpassing 70-75\% of existing models.

\item For GAI, the reduction in operation costs and the performance improvements receive more attention. For example, GLaM~\cite{du2022glam} activates only 8\% of its parameters for each input, consuming only one-third of the energy and requiring half the computational flops for inference compared to GPT-3.
\end{itemize}

While the operation cost of GAI models is reduced, the overall resources required for GAI remain substantial as the model contains a large number of parameters in total. Therefore, utilizing devices in mobile edge networks to support MoE-enhanced GAI models is imperative yet challenging.

\section{MoE Enhanced Content Generation Supported by Mobile Edge Networks}
In this section, we discuss how mobile edge networks support GAI model with MoE structure, including research challenges, the proposed solution, and the case study.
\subsection{Research Challenges}
Taking the AIGC service as an example, when the user initiates a request, the model on user devices evaluates the task and local computing resources. If resources are insufficient, the user device transfers certain tasks to other mobile edge devices for completion, allowing the mobile edge network to support GAI models with MoE architecture. While holding great potential, there are still key challenges to be addressed.
 
\begin{itemize}
\item \textbf{Time-varying Channel State}: The wireless channel conditions can impact the subtask transferring and generated content uploading. In mobile environments, for instance, the time-varying channel may reduce signal-to-noise ratio (SNR), leading to a higher bit-error rate. This can directly affect accuracy of task transferring, potentially causing detrimental effects on the overall quality of service (QoS).

\item \textbf{Bandwidth Consumption}: Both transferring generation subtasks and uploading generated content consume bandwidth resources. Particularly, when an edge device uploads the generated content to user devices, a larger bandwidth may be required to achieve a lower latency, which leads to considerable costs.

\item \textbf{Computing Resources Consumption}: AIGC service requires substantial computational resources, particularly for complex tasks such as logical reasoning and video generation, which often require larger GAI models. In the MoE architecture, each expert may consume different amounts of computing resources and energy when executing subtasks. Therefore, transferring these subtasks requires consideration of both the transmission and computing capacities of edge networks.

\item \textbf{Incentive Mechanism}: Designing an incentive mechanism is crucial to motivate edge devices to actively participate in content generation. This task requires considering numerous factors, including the computing costs for user and edge devices, communication costs for transferring subtasks and uploading results, the market pricing for per unit of QoS, and so forth.

\item \textbf{Model Upgrade}: The MoE facilitates the incorporation of new experts into GAI models, improving their ability to meet evolving user demands. However, expanding the MoE necessitates retraining the gating network. Moreover, this updated gating network demands a reassessment of the subtasks transfer mechanism to determine if the mobile edge network can accommodate the enhanced model.
\end{itemize}

\subsection{The Proposed Framework}
During content generation, user device's resources may not be sufficient for all selected experts, leading the MoE to omit some subtasks for the overall task completion. For instance, if 8 experts are selected but only 6 can be utilized due to resource constraints, the system needs to exclude 2 experts, relying on the remaining 6. This affects the QoS, particularly if most important experts are left out. To mitigate this, our framework transfers certain subtasks to mobile edge devices, ensuring to complete the task execution. The workflow of our framework is outlined as follows.

\begin{enumerate}
    \item \textbf{Task Decomposition}: When a user initiates a content generation request, the system first analyzes the user's input and divides the task into multiple subtasks. For instance, in text generation tasks, the model can categorize the user's input prompt by word attributes, such as nouns, verbs, and adjectives, thus creating several subtasks.
    
    \item \textbf{Computing Resource Assessment}: Based on the decomposed subtasks, the gating function selects appropriate experts and then assesses the computing resources required to complete the entire content generation task. If the local resources are sufficient, the task is processed locally. Otherwise, some subtasks are transferred to mobile edge devices for computation.

    \item \textbf{Subtask Transferring}: User devices select some edge experts and transfer subtasks to them via wireless networks. Note that each edge device may be situated in a distinct wireless environment and excels in different types of content. Hence, transferring to different experts leads to differences in the content quality and associated costs. 

    \item \textbf{Fianl Results Generation}: Edge devices upload the generated content to the user device, which then combines it with those generated locally to produce the final output and present it to the user.
\end{enumerate}

\begin{figure*}[t]
	\centering
	\includegraphics[width=1\textwidth]{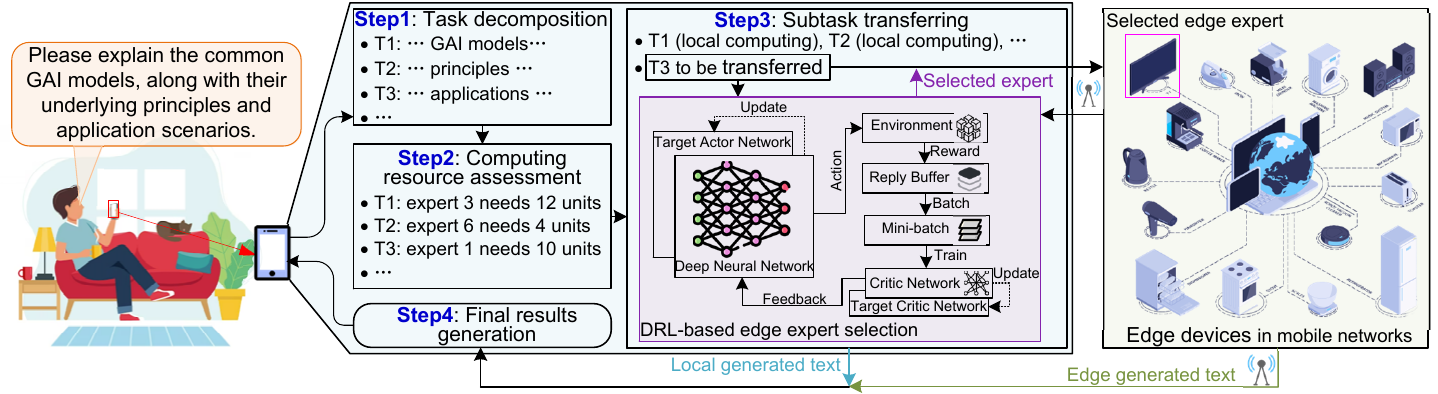}%
	\caption{The operation process of the proposed framework. Upon receiving the prompt, the user device decomposes the task and assesses the computing resources needed for each subtask. Then, considering wireless channel conditions and edge device availability, it selects suitable edge experts and delegates some subtasks to them. Finally, the text generated by the user device is integrated with those from the edge devices to form the final output. }
	\label{framework}
\end{figure*}

Figure~\ref{framework} illustrates the workflow of our proposed framework, which leverages the mobile edge network to ensure the completion of all subtasks, ensuring that the final content fully reflects the user's prompt without missing any tokens. This prevents the need to transfer experts with substantial parameter sizes to the user device, avoiding network congestion. Ideally, the user device transfers subtasks to edge devices that host the most appropriate expert with low communication costs and sufficient computing resources. However, matching experts to subtasks in mobile networks is often uncertain, and wireless conditions differ across edge devices. This causes increased computing and communication costs as a user iteratively seeks for an optimal edge device. Hence, efficiently choosing the most suitable edge device to optimize system utility is a challenging yet crucial issue in our framework.

\subsection{DRL-based Edge Expert Selection}
In the proposed framework, we employ the soft actor-critic (SAC) DRL to tackle the edge expert selection challenge. SAC is grounded in the maximum entropy principle, designed to optimize policy for greater cumulative rewards while focusing on maximizing the trade-off between expected return and entropy. Specifically, regarding the edge expert selection, the definitions of the state space, action space, and reward are as follows.

 \begin{itemize}
\item \textbf{State Space}: The state space consists of two parts. The first is a user-side feature vector, depicting the computing resources required for the subtask, the type of expert network needed, and the communication costs of transferring the subtask. The second part is the edge device feature vector, illustrating the computing resources available, the type of expert network on the edge device, and the corresponding communication costs. Here, the communication cost is linked to various factors, including the SNR, transmitting power, the size of generated content, and other parameters.

\item \textbf{Action Space}: The action space is represented by an integer, indicating which edge device the user device selects to transfer and complete the subtask. Therefore, the actor policy network outputs a logits vector, which is then processed by a softmax operator to determine the probability of selecting each edge device. Finally, the user device selects the edge device with the highest probability and transfers the subtask to it for computation.

\item \textbf{Reward}: The reward is determined by the quality of the final generated content, and the communication and computing cost. The former can be obtained through various evaluation methods, e.g., as given in~\cite{chen2023exploring} and~\cite{du2023enabling}, while the latter includes resources consumed by transferring subtasks and generated content, as well as computing resource consumption.

\end{itemize}

\subsection{Case Study}
\subsubsection{Experimental configuration} Through an example, we demonstrate how to use edge devices to support the MoE based text generation services. We consider a setup with one user device and 30 edge devices. The user initiates a text generation request, and the user user device divides the task into subtasks, transferring one to an edge device for processing. Each edge device has an expert network, skilled in generating text on various topics such as character appearance, landscapes, weather, and so forth. Considering the varying wireless environments of edge devices, the SNR is randomly set between 5 and 20, with the communication bandwidth and transmission power at 1kHz and 0.1W, respectively. Therefore, the communication cost can be calculated as the energy consumed to transfer subtasks and upload generated texts. Meanwhile, the computing cost is defined as a function of the amount of the generated texts. In experiments, the expert networks and gating network are powered by OpenAI's API\footnote{https://openai.com/research/gpt-4}. Although experts at different edge devices specialize in specific areas, they can produce texts beyond their primary areas of expertise. For example, an expert in character descriptions can also generate weather-related content, although the quality might not be as impressive. The final output, an integration of texts from the user device and edge devices, is assessed by the method in~\cite{chen2023exploring}, resulting in a explicit score for the generated text\footnote{https://github.com/HongyangDu/Net4MoE}.
\subsubsection{Performance Analysis}
\begin{figure}[tbp!]
  \centering
  \includegraphics[height=6.5cm]{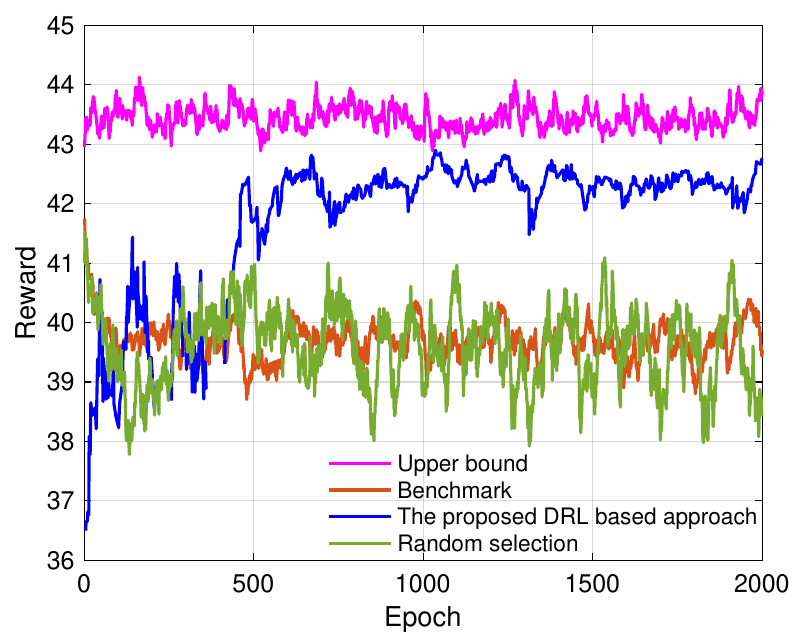} 
  \caption{Reward values versus the number of epoch in DRL.} 
  \label{TR}
\end{figure}

Figure~\ref{TR} illustrates the reward comparison between the proposed selection algorithm and other methods. Here, the upper bound represents the reward obtained when the user device has enough resources to complete all subtasks itself. The benchmark reflects the reward obtained when the user device, due to limited resources, abandons one subtask but completes the rest. The results show that as learning progresses, the edge devices selected by the proposed algorithm gradually optimize, with the corresponding rewards stabilizing at 42.6, close to the upper bound with the reward of 43.6. Meanwhile, our algorithm outperforms the benchmark and random selection methods. This firstly illustrates the necessity of our framework, i.e., transferring subtasks to edge devices is crucial for QoS when user device have insufficient resources. Secondly, it validates the effectiveness of our selection algorithm, which avoids the resource wastage from the user device blindly trying different edge devices.

\begin{figure}[tbp!]
  \centering
  \includegraphics[height=5cm]{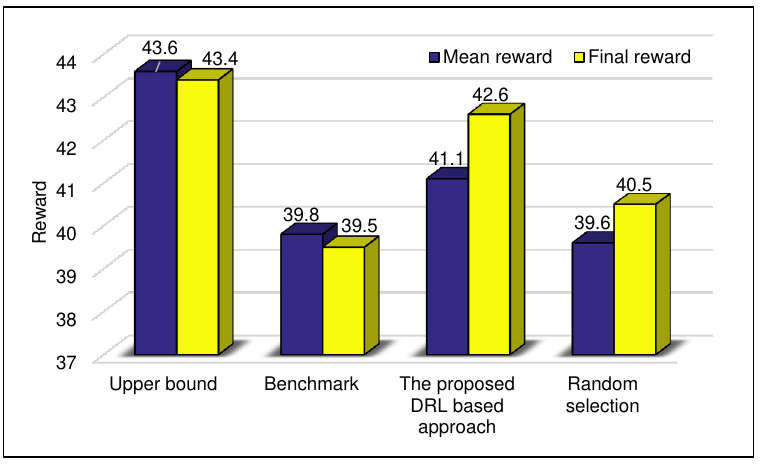} 
  \caption{ Average and final rewards comparison of different methods.} 
  \label{BAR}
\end{figure}

Figure~\ref{BAR} presents the average and final rewards for all methods. As shown, the average and final rewards of the proposed method are 41.1 and 42.6, respectively, surpassing both the benchmark (39.8 and 39.5) and random selection (39.6 and 40.5), demonstrating its effectiveness in selecting suitable experts from the edge network for subtasks. Besides, the rewards for the proposed algorithm are below the upper bound. This is because the experts and the gating network on the user device are trained jointly, enabling more effective teamwork compared to finishing the subtask on the edge device. Therefore, completing all subtasks on the user's device yields better quality text.

\section{Future Directions}
\subsection{Semantic Communications}
Semantic communication focuses on transmitting the semantic information of the source, and hence the transmitter and receiver need to share background information so that the semantic encoding and decoding can be performed effectively~\cite{wang2023semantic}. Within this framework, the MoE can be used to construct the semantic decoder, which can include multiple semantic experts with diverse background knowledge. During the decoding process, the system can perform decoding by activating a subset of semantic experts based on demand, thereby enhancing decoding performance and efficiency.
\subsection{Integrated Sensing and Communications}
In integrated sensing and communications (ISAC)~\cite{wang2023generative}, processing of the received wireless signals enables simultaneous user communication and sensing, such as target detection and tracking. Within this process, MoE can be adopted to build a multi-tiered signal processing architecture, which comprises multiple experts specialized in handling communication and sensing tasks, as well as a gating network. Such architecture can selectively activate certain experts to process the received signals based on different wireless channel conditions and specific sensing tasks, thereby further enhancing the overall performance of ISAC systems.

\subsection{Space-air-ground Integrated Network}
The space-air-ground integrated network (SAGIN), which includes technologies from the ground, air and space networks, is a vital part of 6G for realizing ubiquitous connectivity~\cite{zhang2023generative}. Given its unique cross-layer architecture, it is essential to employ MoE to address the challenges within different layers. For instance, experts responsible for ground-based networks should concentrate on resource allocation in ground-based Internet and mobile communication networks, while experts for space-based networks need to focus on the cooperative transmission among satellites and airborne base stations, to ensure global connectivity.

\section{Conclusions}
This paper proposed the use of mobile edge networks to support the deployment and operation of MoE-based GAI models on devices with limited resources. Specifically, we proposed a framework that transfers certain subtasks of GAI models to edge devices for completion, ensuring the performance when resources of the user devices are limited. Considering the communication, computation, and other related costs during transferring, we further proposed a DRL-based algorithm, which selects suitable edge devices to finish the transferred subtask, thereby ensuring the overall QoS of the model. Experimental results validated the necessity and effectiveness of the proposed framework, providing valuable insights for deploying large GAI models on devices with constrained resources.
\bibliographystyle{IEEEtran}
\bibliography{Ref.bib} 

\end{document}